\documentclass{ws-procs10x7}
\usepackage{balance}
\usepackage{bm}

\newcolumntype{d}[1]{D{.}{.}{#1}}

\makeindex
\begin{document}

\title{Stability of superfluid Fermi gases in optical lattices}

\author{Yoshihiro Yunomae$^*$}

\address{Department of Physics, Waseda University, Okubo, Shinjuku-ku, Tokyo 169-8555, Japan\\$^*$E-mail: yunon@kh.phys.waseda.ac.jp}

\author{Ippei Danshita}

\address{Department of Physics, Faculty of Science, Tokyo University of Science, Kagurazaka, Shinjuku-ku, Tokyo 162-8601, Japan}

\author{Daisuke Yamamoto}

\address{Department of Physics, Waseda University, Okubo, Shinjuku-ku, Tokyo 169-8555, Japan}

\author{Nobuhiko Yokoshi}

\address{Nanotechnology Research Institute, AIST, Tsukuba 305-8568, Japan\\CREST(JST), Saitama 332-0012, Japan}

\author{Shunji Tsuchiya}

\address{Department of Physics, Keio University, Hiyoshi, Kohoku-ku, Yokohama 223-8522, Japan\\CREST(JST), Saitama 332-0012, Japan}

%%%%%%%%%%%%%%%%%%%%%%%%%%%%%%%%%%%%%%%%%%%%%%%%%%%%%%%%%%%%%%%%%%%%%%%%%
% You may repeat \author  \addres as often as necessary                 %
%%%%%%%%%%%%%%%%%%%%%%%%%%%%%%%%%%%%%%%%%%%%%%%%%%%%%%%%%%%%%%%%%%%%%%%%%

\twocolumn[\maketitle 
\abstract{Critical velocities of superfluid Fermi
gases in optical lattices are theoretically investigated across the
BCS-BEC crossover. We calculate the excitation spectra in the presence
of a superfluid flow in one- and two-dimensional optical lattices. 
It is found that the spectrum of low-lying Anderson-Bogoliubov (AB) mode exhibits a
roton-like structure in the short-wavelength region due to the strong charge
density wave fluctuations, and with
increasing the superfluid velocity one of the roton-like minima reaches 
zero before the single-particle spectrum does. This means that superfluid
Fermi gases in optical lattices are destabilized due to spontaneous
emission of the roton-like AB mode instead of due to Cooper pair
breaking.}
]

\section{Introduction}
According to Landau criterion, superfluid flow of bosonic particles
propagates without dissipation, unless its group velocity exceeds a certain
critical value and spontaneous creation of excitations is
induced.\cite{Landau} The direct experimental verifications of the
criterion have been attempted mainly in superfluid $^4$He \cite{tilley}
and dilute atomic Bose gases.\cite{onofrio} These studies have inspired
various discussions and contributed to understanding of the superfluid
phenomenon itself.

The recent realization of the BCS-BEC crossover in ultracold
Fermi gases\cite{Regal} has raised a new question, i.e., how does the critical
velocity for bosonic superfluids change in the BCS regime? Recently,
Miller {\it et al.} investigated critical velocities of superfluid Fermi gases across the
BCS-BEC crossover in a moving one-dimensional (1D) optical lattice.\cite{Miller} 
They determined the critical velocities at which the abrupt decrease of fermionic
pairs in the condensate occurs, when the velocity of the lattice
potential is varied.

Theoretically, critical velocities in the BCS-BEC crossover have been
investigated in a uniform system.\cite{COMB} It was claimed that
the instability of superfluid flow in a uniform system is induced by
Cooper pair breaking in the BCS regime.\cite{COMB}
In this paper, motivated by the experiment done by
Miller {\it et al.},\cite{Miller} we investigate critical velocities of
superfluid Fermi gases in the BCS-BEC crossover in optical lattices.
We calculate excitation spectra and determine critical velocities in
1D and 2D optical lattices.
We find that the excitation spectrum of the low-lying
Anderson-Bogoliubov (AB) mode\cite{COMB,Anderson,Pita} has a
characteristic roton-like structure, and the instability of superfluid
flow in the BCS regime is induced by a different mechanism from that in
a uniform system: the spontaneous emission of AB mode destabilizes the superfluidity.

%%%%%%%%%%%%%%%%%%%%%%%%%%%%%%%%%%%%%%%%%%%%%%%%%%%%%%%%%%%%%%%%%%%%%%%%%%%
%%%%%%%%%%%%%%%%%%%%%%%%%%%%%%%%%%%%%%%%%%%%%%%%%%%%%%%%%%%%%%%%%%%%%%%%%%%
\section{Formalism}
%%%%%%%%%%%%%%%%%%%%%%%%%%%%%%%%%%%%%%%%%%%%%%%%%%%%%%%%%%%%%%%%%%%%%%%%%%%
%%%%%%%%%%%%%%%%%%%%%%%%%%%%%%%%%%%%%%%%%%%%%%%%%%%%%%%%%%%%%%%%%%%%%%%%%%%
We consider two-component atomic Fermi gases with equal
populations. We assume that the lattice potential is sufficiently deep
that the tight-binding approximation is valid. Thus, the system can be
described by a single-band Hubbard model
%%%%%%%%%%%%%%%%%%%%%%%%%%%%%%%%%
\begin{eqnarray}
H
&=&-J\sum_{\langle i,j \rangle, \sigma}
\left(c_{i\sigma}^\dagger c_{j\sigma}+h.c.\right) \nonumber \\
&&+U\sum_i c_{i\uparrow}^\dagger c_{i\downarrow}^\dagger c_{i\downarrow}c_{i\uparrow},
\end{eqnarray} 
%%%%%%%%%%%%%%%%%%%%%%%%%%%%%%%%%
where $c_{j\sigma}$ is an annihilation operator of an atom with mass $m$ and
pseudospin $\sigma=\uparrow,\downarrow$ on the $j$-th site, $J$ is a
hopping energy between nearest-neighbor sites, and $U(<0)$ is an on-site
attractive interaction. We set $\hbar=1$ throughout the paper. 

%When the optical lattices move with a constant velocity $-{\bm v}$,
We assume that a superfluid flows with the velocity $\bm v$ in the
coordinate system fixed with respect to the lattice potential, i.e.
Cooper pairs have the center of mass quasimomentum $\bm q=2m{\bm v}$. The
corresponding order parameter is given by
%%%%%%%%%%%%%%%%%%%%%%%%%%%%%%%%%
\begin{equation}
\Delta_{\bm v}=-\frac{U}{M}\sum_{\bm k}\langle c_{-{\bm k}+m{\bm v}\downarrow}c_{{\bm k}+m{\bm v}\uparrow}\rangle,
\end{equation}
%%%%%%%%%%%%%%%%%%%%%%%%%%%%%%%%%
where $M$ is the number of lattice sites. 
By diagonalizing the Hamiltonian within the BCS mean-field
approximation, we obtain the single-particle excitation energy as
%%%%%%%%%%%%%%%%%%%%%%%%%%%%%%%%%
\begin{equation}
E_{\bm v}^\pm({\bm k})=\eta_{\bm v}(\bm k)
\pm\sqrt{(\varepsilon_{\bm v}(\bm k)-\mu)^2+|\Delta_{\bm v}|^2},
\label{single}
\end{equation}
%%%%%%%%%%%%%%%%%%%%%%%%%%%%%%%%%
where $\eta_{\bm v}(\bm k)=2J \Sigma_{\nu} \sin(m v_{\nu} d)\sin(k_{\nu}
d)$ and $\varepsilon_{\bm v}({\bm k})=2J\Sigma_{\nu}(1-\cos(m v_{\nu}
d)\cos(k_{\nu} d))$ with $d$ being the lattice constant. The energy gap
$\Delta_{\bm v}$ and the chemical potential $\mu$ are determined by
solving the gap equation and the number equation
self-consistently. 
This scheme interpolates the BCS and BEC regimes at $T=0$.\cite{Leggett}

To determine the critical velocities, it is necessary to calculate the
excitation spectra in the presence of a superfluid flow. For neutral
fermionic superfluids, there exists a low-lying AB mode\cite{COMB,Anderson,Pita} in
addition to the single-particle excitation. The spectrum of the AB mode
can be obtained from the pole of the density response function $\chi({\bm
q},\omega).$\cite{Cote} We calculate $\chi({\bm q},\omega)$ by employing the
generalized random phase approximation (GRPA) developed by C$\hat{\rm
o}$t$\acute{\rm e}$ and Griffin.\cite{Cote}
%Here we do not refer to the details of the calculations, which will be presented elsewhere\cite{yunon}.

%%%%%%%%%%%%%%%%%%%%%%%%%%%%%%%%%%%%%%%%%%%%%%%%%%%%%%%%%%%%%%%%%%%%%%%%%%%%%
%%%%%%%%%%%%%%%%%%%%%%%%%%%%%%%%%%%%%%%%%%%%%%%%%%%%%%%%%%%%%%%%%%%%%%%%%%%%%
\section{Results}
%%%%%%%%%%%%%%%%%%%%%%%%%%%%%%%%%%%%%%%%%%%%%%%%%%%%%%%%%%%%%%%%%%%%%%%%%%%%%
%%%%%%%%%%%%%%%%%%%%%%%%%%%%%%%%%%%%%%%%%%%%%%%%%%%%%%%%%%%%%%%%%%%%%%%%%%%%%
Conditions for Cooper pair breaking are determined from the
single-particle excitation energy Eq.~(\ref{single}). When the
absolute value of the superfluid velocity $\bm v$ exceeds a certain
critical value $v_{\rm pb}$, one finds that $E_{\bm v}^+({\bm k})$ becomes
negative. 
The negative energy of the single-particle excitation indicates the occurrence 
of spontaneous Cooper pair breaking. Thus, the critical velocity for
Cooper pair breaking is given by $v_{\rm pb}.$\cite{Miller,COMB,Rodriguez}

%%%%%%%%%%%%%%%%%%%%%%%%%%%%%%%%%%%%%%%%%%%%%%%%%%%%%%%%%%%%%%%%%%%%%%%%%%%%
\begin{figure}[t]
\includegraphics[width=13pc]{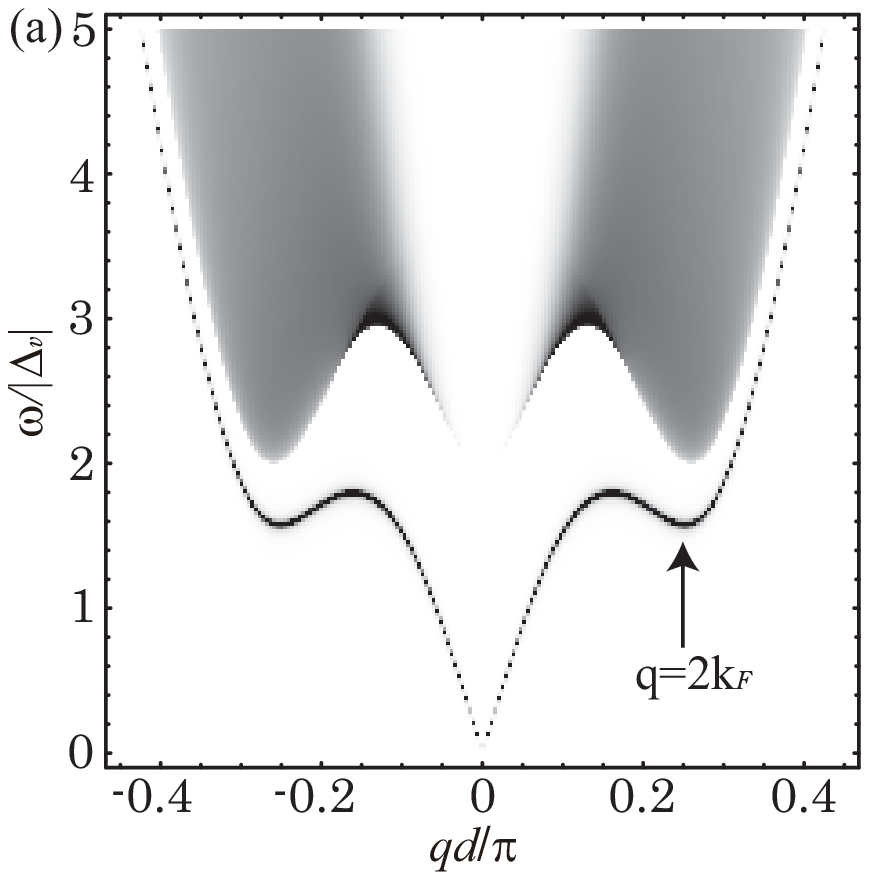}
\includegraphics[width=13pc]{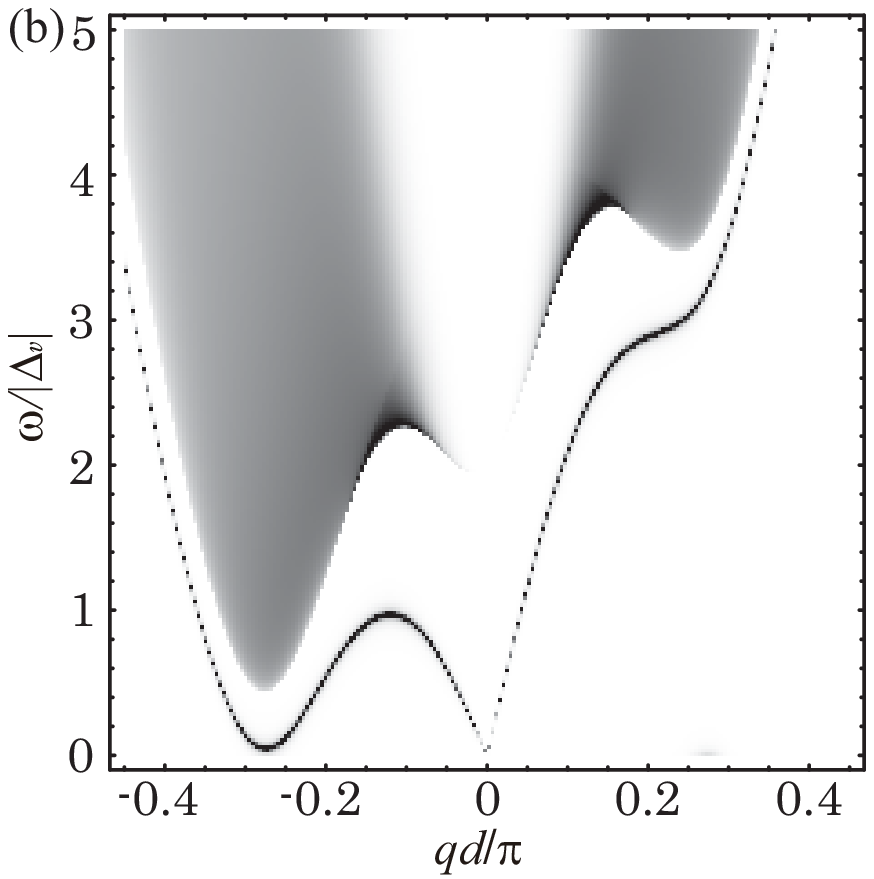}
\caption{
Dynamical structure factors $S(q,\omega)$ in 1D optical lattices are shown for
(a) current-free ($v=0$) and (b) current-carrying cases ($v=0.095/md$),
when $n=0.25$ and $U=-1.0J$.
The dark region indicates large $S(q,\omega)$.
In each panel, the upper gray region corresponds to the particle-hole continuum,
and the lower dashed curve is the spectrum of the AB mode. }
% For the case with zero lattice velocity $v=0$); the gap energy and
% chemical potential are calculated as $|\Delta_v|=0.09863 J$ and
% $\mu=0.16527J$. (b) For the case with near the critical velocity
% $(v=0.095/md)$; $|\Delta_v|=0.09925 J$ and $\mu = 0.17366 J$.} 
\label{nov}
\end{figure}
%%%%%%%%%%%%%%%%%%%%%%%%%%%%%%%%%%%%%%%%%%%%%%%%%%%%%%%%%%%%%%%%%%%%%%%%%%%%
In order to compare the instabilities induced by two kinds of
excitations, i.e., the single-particle excitation and AB mode excitation,
it is necessary to monitor both the excitation spectra in parallel. 
For that purpose, it is useful to plot the dynamical structure foctor $S({\bm
q},\omega)=-{\rm Im}\chi({\bm q},\omega)/\pi$. First, we investigate the
case of 1D optical lattices. In Fig.~\ref{nov}, $S(q,\omega)$ is
shown for (a) current-free and (b) current-carrying cases. 
In Fig.~\ref{nov}, the upper gray regions are the particle-hole
continuums, and the lower boundaries of the
particle-hole continuums correspond to $\omega={\rm min}\left[E_{\bm v}^+(\bm k+\bm q)-E_{\bm v}^-(\bm k)\right]$.
The lower curves in Fig.~\ref{nov} are the spectrum of AB mode, which
becomes phonon-like when $qd\ll 1$.\cite{Anderson}
It should be noted that the spectrum of the AB mode has roton-like
minima at $|q| \simeq 2k_F$ where $k_F$ is the Fermi wavevector. 
Such a roton-like structure also appears in the spectrum of AB mode
in uniform 1D Fermi gases.\cite{ALM}
The energies of the roton-like minima decrease as one approaches the
half-filling when $v=0$. They eventually become zero at $k=\pi/d$ at the
half-filling. This suggests that exciting density fluctuations with $k=\pi/d$
costs zero energy and the superfluid ground state is in competition with the
charge density wave (CDW) state.
Thus, the roton-like minima reflect the strong CDW fluctuations.

%%%%%%%%%%%%%%%%%%%%%%%%%%%%%%%%%%%%%%%%%%%%%%%%%%%%%%%%%%%%%%%%%%%%%%%%%%%%
\begin{figure}[t]
\includegraphics[width=14pc]{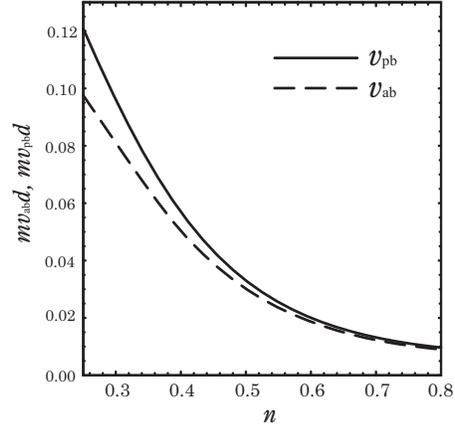}
\caption{
Critical velocities $v_{\rm pb}$ and $v_{\rm ab}$ as functions of the
 filling $n$ when $U=-1.0J$. The solid line represents $v_{\rm pb}$ and
 the dashed line represents $v_{\rm ab}$. 
%One can see that both of them become smaller as $n$ increases to unity
%(half-filling), and $v_{\rm ab}$ is smaller than $v_{\rm pb}$.
} 
\label{cv1D}
\end{figure}
%%%%%%%%%%%%%%%%%%%%%%%%%%%%%%%%%%%%%%%%%%%%%%%%%%%%%%%%%%%%%%%%%%%%%%%%%%%%
In the presence of a superfluid flow with $v>0$, the spectrum slants towards
the left side (see Fig.~\ref{nov}(b)). The critical velocity due to
pair breaking $v_{\rm pb}$ coincides with the velocity at which the
particle-hole continuum touches the line $\omega=0$. 
Since the AB mode spectrum locates below the particle-hole
continuum, one of the roton-like minima reaches zero at the
velocity $v_{\rm ab}$ which is smaller than $v_{\rm pb}$. 
According to
the Landau criterion,\cite{Landau} this means that the spontaneous emission of the AB
mode is induced before the Copper pair breaking occurs. 
In Fig.~\ref{cv1D}, the critical velocities
$v_{\rm pb}$ and $v_{\rm ab}$ are plotted as functions of the filling
$n$. One clearly sees that $v_{\rm ab}$ is always smaller
than $v_{\rm pb}$, and that both $v_{\rm ab}$ and $v_{\rm pb}$ decrease
monotonically with increasing filling. 
The difference between the two velocities diminishes near the
half-filling ($n\simeq 1$).
%, and it becomes difficult to distinguish the instabilities in the two scenarios. 

\begin{figure}[t]
\includegraphics[width=13pc]{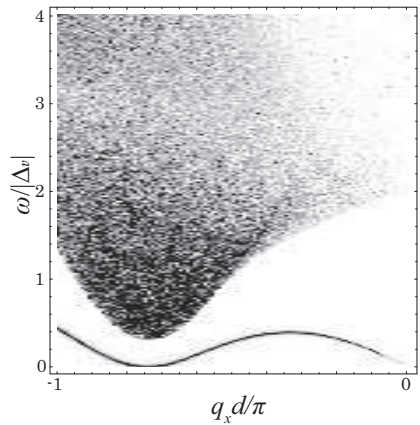}
\caption{Dynamical structure factor $S({\bm q},\omega)$ in 2D optical
 lattices for $q_{x}=q_{y}$. We set $U=-4.5J$, $n=0.5$, and ${\bm
 v}=(v,v)/\sqrt{2}$ with $v=0.467/md$. The dark region indicates large $S(\bm q,\omega)$. The upper gray region corresponds
 to the particle-hole continuum, and the lower curve is the spectrum of
 AB mode.}
\label{2D}
\end{figure}

We proceed to discuss Fermi gases in 2D optical lattices. In Fig.~\ref{2D},
we plot the dynamical structure factor $S(\bm q,\omega)$ for
$q_{x}=q_{y}$, when $\bm v=(v/\sqrt{2},v/\sqrt{2})$.
Similarly to the 1D case, the spectrum of AB mode has a
roton-like minimum in Fig.~\ref{2D}.
It reaches zero before the particle-hole continuum does when
$v$ is increased. Thus, $v_{\rm ab}$ is smaller than
$v_{\rm pb}$, and the instability induced by emission of the AB mode
also dominates in 2D optical lattices.
Furthermore, it is found that the direction of the superfluid velocity
$\bm v$ affects the spectrum of AB mode.
The energy of the roton-like minimum decreases more rapidly for ${\bm
v}=(v/\sqrt{2},v/\sqrt{2})$ than for ${\bm v}=(v,0)$.
This implies that the CDW fluctuation is
enhanced when the supercurrent is flowing in the direction of $(\pi,\pi)$.
%This implies that the nesting effect of
%the Fermi line for 2D square lattice plays an important role. 
%However, the difference between ${\bm v}_{\rm ab}$ and ${\bm v}_{\rm pb}$ is
%smaller compared with the one in 1D case. In order to distinguish them
%in experiment, relatively strong attraction is needed.  

%%%%%%%%%%%%%%%%%%%%%%%%%%%%%%%%%%%%%%%%%%%%%%%%%%%%%%%%%%%%%%%%%%%%%%%%%%%%
%%%%%%%%%%%%%%%%%%%%%%%%%%%%%%%%%%%%%%%%%%%%%%%%%%%%%%%%%%%%%%%%%%%%%%%%%%%%
%%%%%%%%%%%%%%%%%%%%%%%%%%%%%%%%%%%%%%%%%%%%%%%%%%%%%%%%%%%%%%%%%%%%%%%%%%%%%
%%%%%%%%%%%%%%%%%%%%%%%%%%%%%%%%%%%%%%%%%%%%%%%%%%%%%%%%%%%%%%%%%%%%%%%%%%%%%
\section{Conclusion}
%%%%%%%%%%%%%%%%%%%%%%%%%%%%%%%%%%%%%%%%%%%%%%%%%%%%%%%%%%%%%%%%%%%%%%%%%%%%%
%%%%%%%%%%%%%%%%%%%%%%%%%%%%%%%%%%%%%%%%%%%%%%%%%%%%%%%%%%%%%%%%%%%%%%%%%%%%%
In summary, we have studied critical velocities of superfluid Fermi
gases in 1D and 2D optical lattices using the attractive Hubbard model. 
We calculated the excitation spectra of single-particle excitation
and AB mode by employing the GRPA.
We have shown that the spectrum of AB mode has a roton-like structure at $|q|\simeq
2k_F$, and it always lies below the particle-hole continuum due to the
strong CDW fluctuations in optical lattices.
As the velocity of the superfluid flow is increased, the energy of a
roton-like minimum decreases and the instability due to the emission of
roton-like excitations was shown to occur before Cooper pair breaking occurs.
We note that Burkov and Paramekanti have proposed another type of current-induced instability of superfluid Fermi gases in optical lattices.\cite{Burkov}

%As for the case in three-dimensional (3D) lattices, Koponen {\it et al.}
%calculated the AB mode with no superfluid flow and showed that it lies
%below the particle-hole continuum\cite{Koponen}. Since the level
%repulsion always acts between the continuum and the AB mode, the
%instability of our scenario is expected to occur also in higher
%dimension. Moreover, given that recent experiments regarding Bose gases
%loaded into 3D optical lattices have measured the superfluid critical
%velocity\cite{Mun}, the stability of Fermi gases in moving 3D lattices
%can be studied in future experiments.  

%%%%%%%%%%%%%%%%%%%%%%%%%%%%%%%%%%%%%%%%%%%%%%%%%%%%%%%%%%%%%%%%%%%%%%%%%%%%%
%%%%%%%%%%%%%%%%%%%%%%%%%%%%%%%%%%%%%%%%%%%%%%%%%%%%%%%%%%%%%%%%%%%%%%%%%%%%%
\section*{Acknowledgments}
%%%%%%%%%%%%%%%%%%%%%%%%%%%%%%%%%%%%%%%%%%%%%%%%%%%%%%%%%%%%%%%%%%%%%%%%%%%%%
%%%%%%%%%%%%%%%%%%%%%%%%%%%%%%%%%%%%%%%%%%%%%%%%%%%%%%%%%%%%%%%%%%%%%%%%%%%%%
We would like to thank K. Kamide, S. Kurihara, and T. Nikuni for
valuable comments and discussions. I.D. and D.Y. are supported by a
Grant-in-Aid from JSPS.

\end{document}